# Quenched Heavy-Light Decay Constants

UKQCD Collaboration – presented by Hartmut Wittig

Physics Department, The University, Southampton SO9 5NH, UK

We present decay constants from a quenched simulation using both propagating quarks and the static theory at $\beta = 6.2$ on $24^3 \times 48$. The results using propagating quarks are used to test scaling laws predicted by the heavy quark symmetry.

## 1. PRELIMINARIES

We present results for heavy-light decay constants for both propagating heavy quarks and the static theory [1], using the $O(a)$-improved Sheikholeslami-Wohlert (SW) action [2]. The simulations were done on a $24^3 \times 48$ lattice at $\beta = 6.2$ [4,5]. Results for propagating heavy quarks were obtained from 60 configurations, whereas results for the static theory were calculated for a subset of 20 configurations. The SW action $S_F^{SW}$ is obtained from the usual Wilson action $S_F^W$ via

$$S_F^{SW} = S_F^W - i\frac{\kappa}{2} \sum_{x,\mu,\nu} \overline{q}(x)\, F_{\mu\nu}(x)\sigma_{\mu\nu}\, q(x). \quad (1)$$

$O(a)$-improved matrix elements involving a light quark $q(x)$, a heavy quark $Q(x)$ and some Dirac matrix $\Gamma$ are obtained using "rotated" operators

$$\overline{Q}(x)\bigl(1 + \tfrac{1}{2}\overleftarrow{\slashed{D}}\bigr)\, \Gamma\, \bigl(1 - \tfrac{1}{2}\overrightarrow{\slashed{D}}\bigr) q(x) \quad (2)$$

in the conventional approach [3]. In the static theory the rotation is only performed on the light quark field $q(x)$.

Heavy-light meson propagators were calculated from quark propagators at three values of the hopping parameter of the light quark ($\kappa_l = 0.14144, 0.14226, 0.14262$) and at four values for the heavy quark ($\kappa_h = 0.121, 0.125, 0.129, 0.133$). According to refs. [4,5], the chiral limit is reached at $\kappa_{\rm crit} = 0.14315(2)$, the strange quark is at $\kappa_s = 0.1419(1)$, and the charm quark roughly corresponds to $\kappa_{\rm charm} \simeq 0.129$.

The renormalisation constants $Z_A$, $Z_V$, relating the lattice axial and vector currents to their continuum counterparts were defined from their perturbatively determined values for the SW action. Using a "boosted" gauge coupling $g_{\rm eff}^2$ we find $Z_A = 1 - 0.02\, g_{\rm eff}^2 \simeq 0.97$ and $Z_V \simeq 0.83$ for propagating heavy quarks. When matching the static lattice theory to the full theory at scale $m_b$ one obtains [6]

$$Z_A^{\rm stat} \simeq 0.79. \quad (3)$$

For the inverse lattice spacing in physical units we take a value of 2.7(1) GeV which we obtain from both the string tension $a\sqrt{K}$ and $am_\rho$ [4]. However, using $af_\pi$ to set the scale results in a very high value, namely $a^{-1} = 3.4\,{\rm GeV}$. Therefore we use

$$a^{-1} = 2.7\; {}^{+\,7}_{-\,1}\; {\rm GeV} \quad (4)$$

and treat the error on $a^{-1}$ as a systematic error on all results to follow.

## 2. PROPAGATING HEAVY QUARKS

The pseudoscalar decay constant $f_P$ is obtained from fitting the ratio

$$\frac{\sum_{\vec{x}} \langle A_4^L(x) P^S(0)\rangle}{\sum_{\vec{x}} \langle P^S(x) P^S(0)\rangle} \to \frac{f_P M_P}{Z_A Z_{P^S}} \tanh M_P(\tfrac{T}{2} - t) \quad (5)$$

for $t = 15, \ldots, 22$. Here, $P^S$ denotes a smeared pseudoscalar current, and the wavefunction $Z_{P^S} = \langle 0 | P^S | P \rangle$ and mass $M_P$ are determined from a separate fit to the smeared-smeared pseudoscalar correlator for $t = 13, \ldots, 22$.

The fundamental scaling law for the quantity $f_P \sqrt{M_P}$ reads

$$f_P \sqrt{M_P} = {\rm const.} \times \bigl[\alpha_s(M_P)\bigr]^{-2/\beta_0}. \quad (6)$$

In order to cancel the residual logarithmic dependence on $M_P$ we define the scaling quantity

$$\Phi(M_P) \equiv Z_A^{-1} f_P \sqrt{M_P}\, (\alpha_s(M_P)/\alpha_s(M_B))^{2/\beta_0} \quad (7)$$

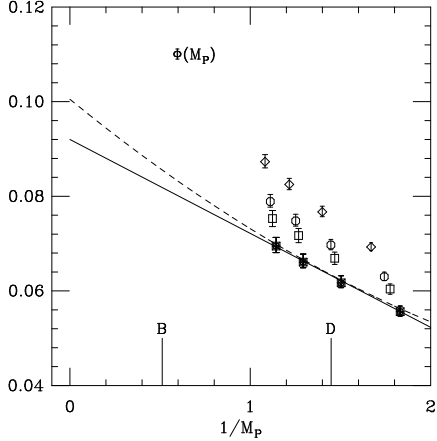

Figure 1. $\Phi(M_P)$ plotted versus $1/M_P$ in lattice units.

We parametrise $\Phi(M_P)$ according to

$$\Phi(M_P) = A\left(1 - \frac{B}{M_P} + \frac{C}{M_P^2}\right) \qquad (8)$$

in order to detect deviations from the scaling law eq. (6).

Fig. 1 shows the fit of the chirally extrapolated values to eq. (8) (dashed line) together with a linear fit with the constraint $C = 0$ (solid line). The linear slope gives the now well-established result that $1/M_P$-corrections amount to about 10% at $M_B$ and about 30% at $M_D$. Ignoring the logarithmic corrections makes the slope slightly more pronounced, however they can by no means account for the observed size of $1/M_P$-corrections.

As is evident from Fig. 1 there is an additional uncertainty in $f_B$ from using either linear or quadratic extrapolations. This uncertainty is added to the systematic error on $f_B$. Our best results at $\kappa_l = \kappa_{\rm crit}$ are thus

$$f_D = 185 \begin{array}{c}+\ 4\\-\ 3\end{array} \begin{array}{c}+\ 42\\-\ 7\end{array} \text{ MeV} \qquad (9)$$

$$f_B = 160 \begin{array}{c}+\ 6\\-\ 6\end{array} \begin{array}{c}+\ 53\\-\ 19\end{array} \text{ MeV} \qquad (10)$$

where the first error is the statistical and the second is the systematic uncertainty. Extrapolating $\kappa_l$ to $\kappa_s$ we find

$$f_{D_s} = 212 \begin{array}{c}+\ 4\\-\ 4\end{array} \begin{array}{c}+\ 46\\-\ 7\end{array} \text{ MeV} \qquad (11)$$

Table 1
Extrapolations of $\widetilde{U}(M)$.

| $M$ | linear | quadratic |
|---|---|---|
| $\infty$ | $1.02\ ^{+\ 5}_{-\ 4}$ | $1.09\ ^{+\ 7}_{-\ 8}$ |
| $(M_B + 3M_{B^*})/4$ | $0.93\ ^{+\ 4}_{-\ 3}$ | $0.96\ ^{+\ 4}_{-\ 5}$ |
| $(M_D + 3M_{D^*})/4$ | $0.77\ ^{+\ 2}_{-\ 2}$ | $0.77\ ^{+\ 2}_{-\ 2}$ |

$$f_{B_s} = 194 \begin{array}{c}+\ 6\\-\ 5\end{array} \begin{array}{c}+\ 62\\-\ 17\end{array} \text{ MeV} \qquad (12)$$

It should be added that our value for $f_{D_s}$ compares well with a recent experimental measurement [7].

The vector decay constant $f_V$ is defined by

$$\langle 0|V_\mu^L|0\rangle \equiv \varepsilon_\mu \frac{M_V^2}{Z_V\,f_V} \qquad (13)$$

which we obtain from a similar ratio as in eq. (5). The spin symmetry of the Heavy Quark Theory predicts the degeneracy of the pseudoscalar and vector decay constants (up to radiative and $1/M$ corrections) in the infinite mass limit

$$\frac{f_V\,f_P}{M} = 1 + \frac{8}{3}\frac{\alpha_s(M)}{4\pi} + O(1/M) \qquad (14)$$

where $M$ is the spin-averaged mass, $M = (M_P + 3M_V)/4$. Defining

$$\widetilde{U}(M) \equiv \frac{f_P\,f_V}{M} \left/ \left(1 + \frac{8}{3}\frac{\alpha_s(M)}{4\pi}\right)\right. \qquad (15)$$

and extrapolating in $1/M$ either linearly or quadratically we should observe $\widetilde{U}(M) \to 1$ as $M \to \infty$. Our results listed in table 1 show indeed that $\widetilde{U}(M)$ satisfies the scaling law in the infinite mass limit which in turn supports our parametrisations of the non-scaling behaviour of $f_P\sqrt{M_P}$.

## 3. STATIC HEAVY QUARKS

For the static theory smeared operators are crucial for the isolation of the ground state. In this simulation we use gauge invariant Jacobi smearing (for details see ref. [8]), and we quote our best result for $N = 140$ iterations in the smearing algorithm. We obtain $Z^L \equiv Z_A^{\rm stat\,-1} f_B\sqrt{M_B/2}$ from fitting

$$\frac{\sum_{\vec{x}}\langle A_4^L(x)A_4^S(0)\rangle}{\sum_{\vec{x}}\langle A_4^S(x)A_4^S(0)\rangle} \to \frac{Z^L\,Z^S}{(Z^S)^2} \qquad (16)$$

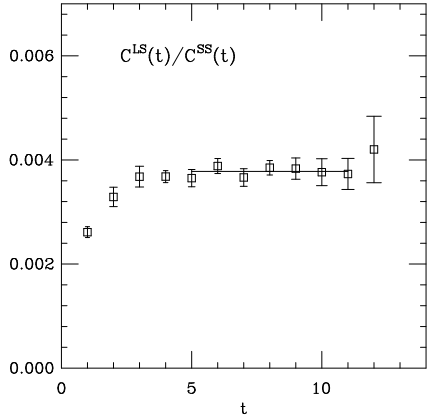

Figure 2. The ratio of correlators of eq. (16).

Table 2
$Z^L \equiv Z_A^{\text{stat}\,-1} f_B \sqrt{M_B/2}$ for different $\kappa_l$.

| $\kappa_l$ | $Z^L$, $N=110$ | $Z^L$, $N=140$ |
|---|---|---|
| 0.14144 | 0.149 $^{+\,7}_{-\,7}$ | 0.142 $^{+\,7}_{-\,6}$ |
| $\kappa_s$ | 0.140 $^{+\,7}_{-\,6}$ | 0.134 $^{+\,7}_{-\,6}$ |
| 0.14226 | 0.137 $^{+\,7}_{-\,6}$ | 0.130 $^{+\,7}_{-\,6}$ |
| 0.14262 | 0.131 $^{+\,7}_{-\,6}$ | 0.125 $^{+\,7}_{-\,7}$ |
| $\kappa_{\text{crit}}$ | 0.124 $^{+\,8}_{-\,7}$ | 0.117 $^{+\,7}_{-\,7}$ |

with $Z^S$ determined from a separate fit to the smeared-smeared correlator. In both cases we fit from $t = 5, \ldots, 11$. In Fig. 2 we show our signal for the ratio of eq. (16).

The results for $Z^L$ are shown in table 2 for $N = 110, 140$, the three values of $\kappa_l$ and for the extrapolations to $\kappa_{\text{crit}}$ and $\kappa_s$. In the chiral limit we obtain $Z^L \simeq 0.12$ which compares to $Z^L = 0.125(8)$ from ref. [9] at $\beta = 6.26$ and $Z^L \simeq 0.10(1)$ from ref. [10] at $\beta = 6.3$. Inserting our values for $a^{-1}$ and $Z_A^{\text{stat}}$ we find for $N = 140$

$$f_B^{\text{stat}} = 253 \,^{+\,16}_{-\,15}\,^{+\,105}_{-\,14} \text{ MeV} \qquad (17)$$

$$f_{B_s}^{\text{stat}}/f_B^{\text{stat}} = 1.14 \,^{+\,4}_{-\,3}. \qquad (18)$$

## 4. CONCLUSIONS

We have demonstrated that the $O(a)$-improved SW action yields consistent results for heavy-light decay constants. The size of $1/M_P$ corrections to the scaling law for $f_P \sqrt{M_P}$ is confirmed. In the static theory we find $f_B \simeq 250$ MeV in agreement with other simulations. We aim at a combined static and conventional analysis which will be possible after the full set of 60 configurations has been calculated for the static case. Results at $\beta = 6.0$ [5] have already shown that a correlated fit to both static and propagating results increases the estimate for $f_B$ by about 20 MeV. If this behaviour persists for our simulations at $\beta = 6.2$, the final answer from the combined analysis will be $f_B \simeq 180$ MeV.

This work was carried out on a Meiko i860 Computing Surface supported by SERC grant GR/G32779, Meiko Limited, and the University of Edinburgh. I thank my colleagues from the UKQCD Collaboration for fruitful discussions.